\documentclass[aps,prmaterials,reprint,amsmath,amssymb,showkeys,superscriptaddress]{revtex4-1}
\pdfoutput=1
\usepackage[english]{babel}
\usepackage[utf8]{inputenc}
\usepackage{graphicx}
\usepackage{xcolor}
\usepackage{physics}
\usepackage[colorinlistoftodos, color=green!40, prependcaption]{todonotes}
\usepackage[pdftex, pdftitle={Article}, pdfauthor={Author}]{hyperref} 
\begin{document}
\title{Electric Field Driven Domain Wall Dynamics in BaTiO\textsubscript{3} Nanoparticles}
\author{Jialun Liu}
\affiliation{London Centre for Nanotechnology, University College London, London WC1E 6BT, United Kingdom}

\author{David Yang}
\affiliation{\hbox{Condensed Matter Physics and Materials Science Department, Brookhaven National Laboratory, Upton, New York 11973, USA}}

\author{Ana F. Suzana}
\affiliation{Advanced Photon Source, Argonne National Laboratory, Lemont, Illinois 60439, USA}

\author{Steven J. Leake}
\affiliation{ESRF, The European Synchrotron, 71 Avenue des Martyrs, CS40220, 38043 Grenoble Cedex 9, France}

\author{Ian K. Robinson}
\affiliation{London Centre for Nanotechnology, University College London, London WC1E 6BT, United Kingdom}
\affiliation{\hbox{Condensed Matter Physics and Materials Science Department, Brookhaven National Laboratory, Upton, New York 11973, USA}}

\date{\today} 

\begin{abstract}
We report a detailed investigation into the response of single BaTiO\textsubscript{3} (BTO) nanocrystals under applied electric fields (E-field) using Bragg Coherent Diffraction Imaging (BCDI). Our study reveals pronounced domain wall migration and expansion of a sample measure in situ under applied electric field. The changes are most prominent at the surface of the nanocrystal, where the lack of external strain allows greater domain wall mobility. The observed domain shifts are correlated to the strength and orientation of the applied E-field, following a side-by-side domain model from Suzana et al. Notably, we identified a critical voltage threshold at +10 V, which leads to irreversible structural changes, suggesting plastic deformation. The findings highlight how surface effects and intrinsic defects contribute to the enhanced dielectric properties of BTO at the nanoscale, in contrast to bulk materials, where strain limits domain mobility. These findings deepen our understanding of nanoscale dielectric behaviour and inform the design of advanced nanoelectronic devices.
\end{abstract}

\maketitle

\section{Introduction} \label{sec:outline}
In ferroelectric materials like BTO, the electric polarisation is inherently linked to the crystal structure where the titanium (Ti\textsuperscript{4+}) ion is displaced from the unit cell centre, creating a dipole along the c-axis in the room-temperature tetragonal phase. For example, nanosized BTO particles show a significant enhancement of their dielectric constant, sufficient for widespread commercial applications in multilayer ceramic capacitors (MLCC) \cite{Curecheriu2010}. Wada et al. \cite{Wada2003} performed powder dielectric measurements on various BTO particles ranging from 17 to 100 nm. Their study showed the BTO particle with a size of 70 nm has the best dielectric constant, around 15000, meaning it is ideal for storing electric energy. Our study examine the dielectric response of hydrothermally synthesised BaTiO\textsubscript{3} nanoparticles, particularly focusing on their behaviour under varying electric fields and domain configurations.

The fundamental nature of the dielectric response in materials involves the displacement of charges proportional to an applied electric field. In ferroelectrics, this response is further complicated by the permanent polarisation that is not only dependent on the applied field but also intimately tied to the structural phase of the material. BTO, for instance, typically exists in a high symmetric (non-ferroelectric) cubic phase at high temperatures but transforms into a low symmetric tetragonal phase when the temperature drops below the Curie temperature around 398 $K$ \cite{Comes1970, Zhong1994}.

\begin{figure*}[ht] 
    \centering 
    \includegraphics[scale=0.12]{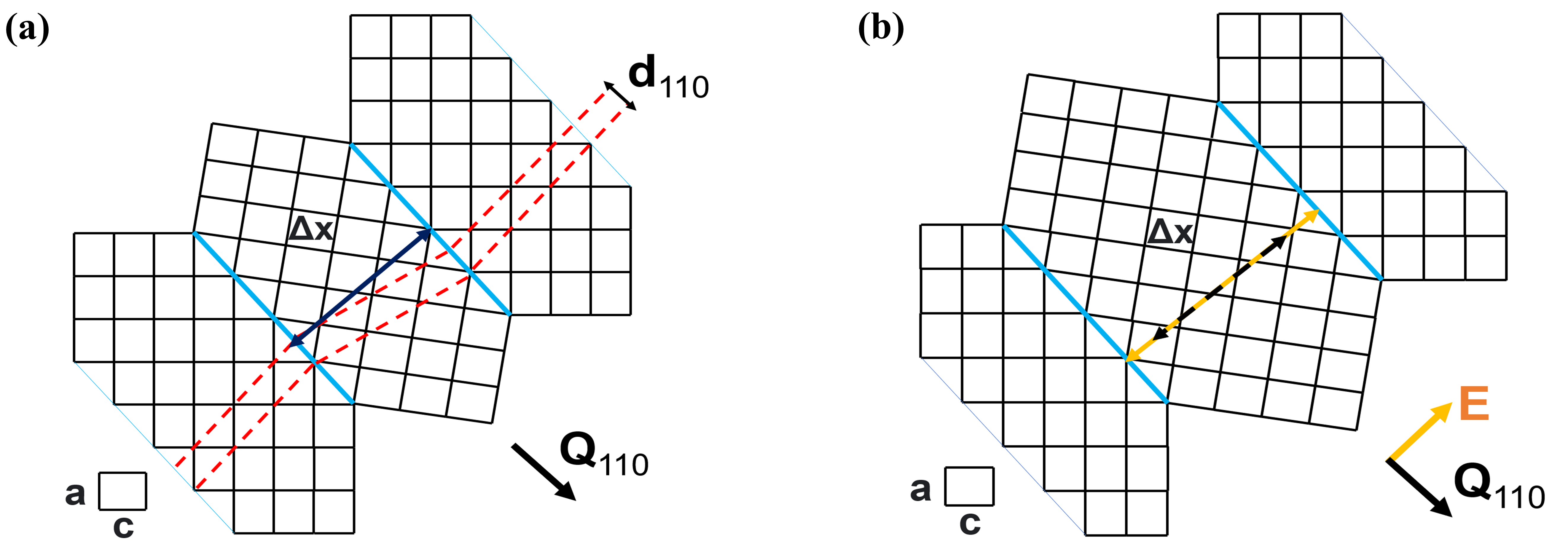}
    \caption{(a) A sketch of the side-by-side model; (b) A model-based prediction of the domain wall movement when an external electric field is applied to a BTO nanoparticle.}
    \label{BTO rec}
\end{figure*}

Models have been proposed to elucidate the enhanced dielectric behaviour observed in nano-scale BTO. These models include, (1) A sudden switch of polarisation within domains, such as a 90-degree rotation, which reflects a dramatic reorientation of the polarisation direction \cite{Mimura2016}. (2) Modulation of the polarisation direction by altering the displacement of the Ti\textsuperscript{4+} ion within the crystal lattice, potentially enhancing or suppressing the overall polarisation \cite{Rudel2020}. (3) The migration of domain walls, which could be influenced by external factors such as the application of an electric field or mechanical stress, leading to configurations like side-by-side domain arrangements as proposed by Suzana et al. \cite{Suzana2023, Diao2020} or in core-shell configurations \cite{Hoshina2008, Hoshina2013}.

Suzana et al. \cite{Suzana2023} introduced a side-by-side model as illustrated in Figure \ref{BTO rec} (a),  detailing the structural characteristics of hydrothermally-synthesised BTO nanoparticles. Comprehensive X-ray diffraction (XRD) analyses found that specific batches of BaTiO\textsubscript{3} samples displayed a mixture of both tetragonal and cubic phases \cite{Suzana2023}. Interestingly, it emerged that the cubic phase was the one that demonstrates a higher dielectric constant in these samples. A comparative analysis between cubic and tetragonal preparations of similar particle size revealed a notable discrepancy: cubic particles exhibited a considerably higher degree of “microstrain" in the Williamson-Hall analysis \cite{Williamson1953} of their powder diffraction patterns. Moreover, the examination of local structure using X-ray pair distribution function (PDF) analysis \cite{Kwei1995, Senn2016} revealed a lower symmetry consistent with tetragonal arrangements \cite{Suzana2023} . Cubic BTO should not be stable at room temperature, but appears to be stabilised in the granular material by mutual strain between nano-domains with tetragonal structure.

Bragg coherent diffraction imaging (BCDI) \cite{Robinson2001,Robinson2006} is a powerful technique for probing strain within single nanocrystals. When a crystal is sufficiently smaller than the coherent beam \cite{Clark2012}, typically less than $1 \mu\text{m} \times 1 \mu\text{m} \times 1 \mu\text{m}$, BCDI can generate high-resolution 3D maps of complex electron density. The amplitude of numbers reflects the “Bragg density", indicating the strength of diffraction for a specific Bragg peak, while the phase represents the projection of local lattice displacement along the Bragg peak direction \cite{Diao2017}.
The phase $\psi_{hkl}$, associated with the crystal displacement vector $\boldsymbol{u}$, is given by the relation \cite{Robinson2009}:
\begin{equation}
    \psi_{hkl}(\boldsymbol{r})=\boldsymbol{Q}_{hkl} \cdot \boldsymbol{u}(\boldsymbol{r})
    \label{eqn:phase}
\end{equation}
where $\boldsymbol{Q}_{hkl}$ is the momentum transfer vector of the Bragg peak. The BCDI image and the diffraction pattern are related by a 3D Fourier transform. The diffraction pattern is an array of complex numbers, of which the phase information cannot be directly measured because pixelated detectors only record the magnitude of the diffraction pattern, giving rise to the famous “phase problem” \cite{Robinson2006, Williams2003}.

Despite this challenge, the phase of the diffraction pattern is encoded within the diffraction fringes, and iterative phase retrieval algorithms have been developed to recover it \cite{Gerchberg1972, Fienup1982}. These algorithms apply constraints in both real and reciprocal space to refine the phase information \cite{Fienup1978, Fienup1986, Yang2022}. A prerequisite for successful phase retrieval is oversampling. According to Bates (1982) \cite{Bates1982}, the fringe spacing of the diffraction pattern must exceed twice the Nyquist frequency, in line with Sayre's interpretation \cite{Sayre1952} of Shannon's information theorem \cite{Shannon1948}.

Remarkably, when imaged using BCDI, it was evident that individual BTO nanoparticles invariably contained multiple domains, each displaying spatially varying strain \cite{Suzana2023}. This led to the model-based prediction as shown in Figure \ref{BTO rec} (b), linking the domain structure within BTO nanoparticles to the dielectric enhancement of BTO nanocrystals. When exposed to an electric field, these domains would undergo a growth and shrinkage response, rather than reorientation of the domain c-axes. This energy-efficient response requires no ionic diffusion, so the behaviour might contribute to the extended service lifetimes and superior high-frequency performance observed in commercial nano-BTO MLCCs. Our investigation here employs \textit{operando} electric field BCDI measurements which confirm this model.

\section{EXPERIMENTAL METHODS} \label{sec:develop}
\subsection{Hydrothermal Synthesis}
We synthesised high-quality nanocrystals, approximately 200 nm in size, using hydrothermal methods \cite{Suzana2023}.These samples were prepared in an autoclave at 240 °C nominal temperature from starting solutions of TiCl\textsubscript{4} and BaCl\textsubscript{2}. They were stirred and mixed in a 50 mL 2 neck round bottom flask. NaOH in 10 mL of deionised water was then added to the system. The whole growth process took place in an autoclave for 48h. The detailed procedure was described by Suzana et al \cite{Suzana2023}.

\begin{figure*}[ht] 
    \centering 
    \includegraphics[scale=0.15]{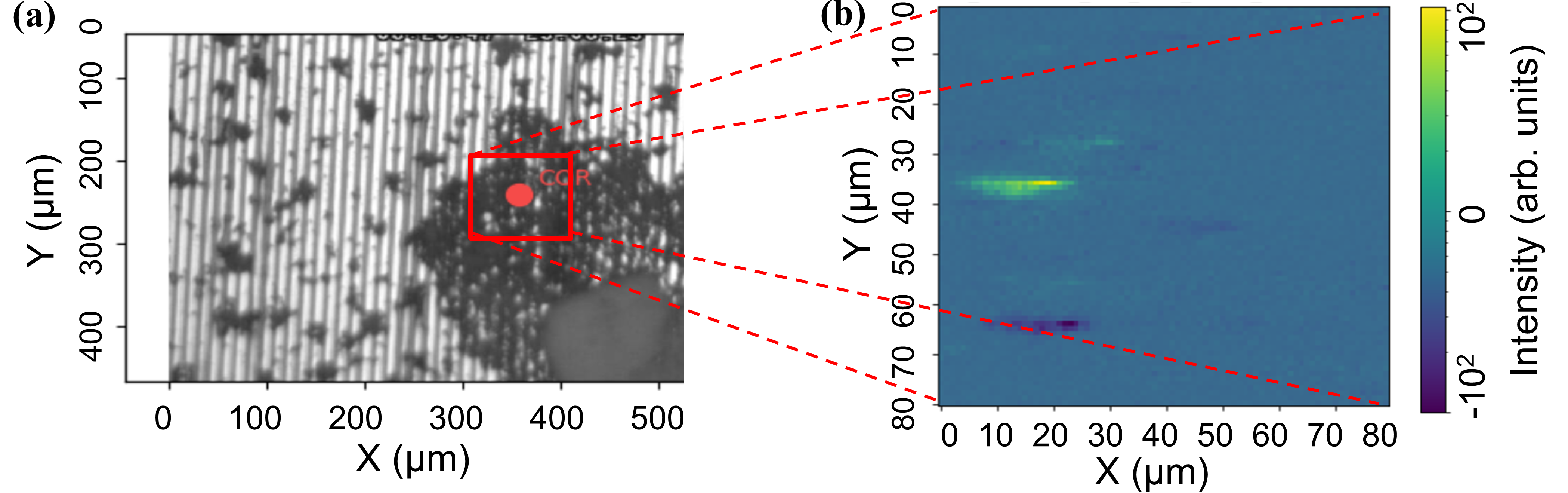}
    \caption[The caption that goes into the list of figures]{(a) Electrode region selected for crystal searching. The beam is passing from left to right, so the Q-vector of the measured particle is roughly parallel to the electrodes (shown as vertical white lines) and perpendicular to the applied electric field. (b) differential k-map lighting up crystals that respond to the change of field. 
 }
    \label{Electrode and k-map}
\end{figure*}

Predominantly, these BTO 240 °C crystals were found to be in the cubic phase, understood to be a strained composite of tetragonal domains \cite{Suzana2023}. We then deposited these nanocrystals onto gold interdigital electrode arrays with 3-micrometer spacing, following the methodology of Kim et al. \cite{Kim2004}. Particles were drop-cast and seamlessly bonded into a continuous film using Tetraethyl orthosilicate (TEOS), followed by a process of annealing \cite{Monteforte2016}. This method was developed to stabilise the particles for BCDI experiments, while preventing dielectric leakage and ensured a uniform electric field throughout the dielectric particles. Examples of the samples used are shown in Fig \ref{Electrode and k-map} (a).
\subsection{Scanning X-ray Diffraction Microscopy (SXDM)} 
In Figure \ref{Electrode and k-map} (a), these electrodes, arranged vertically and perpendicular to the x-ray beam that came from 001 direction, and generated electric field in the z direction. The alignment microscopes at ID-01 enabled the selection of “crowded" regions of particles which bridge the electrodes without gaps, so the full field was seen by the target crystal. By selecting densely clustered BTO particles for the measurement, we minimised capacitance loss due to air exposure and enhanced the electric field response between domain walls. SXDM was then employed to scan locally for crystals that responded to changes in the applied electric field. As shown in Figure \ref{Electrode and k-map} (a), this method divided the confocal microscope field of view into an 80 $\mu$m by 80 $\mu$m region (red box). A raster scan was performed from the top left to the bottom right of the box. The detector was placed according to the calculated [110] 2$\theta$ position, and picked up intensity once a crystal was illuminated during the raster. SXDM was then repeated with an applied electric field. Crystals that responded to the electric field exhibited intensity changes, while those that did not remained neutral. Figure \ref{Electrode and k-map} (b) displays the net intensity differences before and after applying the electric field, with variations between $+/-$ 10V direct current (DC) indicated by positive and negative colours. 

\subsection{BCDI Experiment}
By applying safe voltages up to $+/-$ 10V DC, we could achieve fields up to 3MV/m. This allowed us to investigate the rearrangements of internal domain wall structures using BCDI at beamline ID-01 of the European Synchrotron Radiation Facility (ESRF), utilising x-rays with an energy of 10.5 keV  ($\lambda$ $=$ 0.118 nm). The sample was mounted on a 3-axis goniometer with the Maxipix detector (55 $\mu\text{m}$ $\times$ 55 $\mu\text{m}$ pixels) placed 0.586 m away. 3D BCDI data were collected as rocking curves of detector frames with a step size of $0.01^\circ$ with a 1 s exposure time. The voltage scans are measured in (0 V, 10 V, -5 V, 5 V, -10 V, 0 V) sequence to eliminate any systematic uncertainties. We ensured reproducibility by measuring each crystal twice before changing voltage levels. To account for potential crystal motion during measurements, we employed sub-pixel cross correlation \cite{Guizar2008} processes for pattern alignment. This method provided a more accurate interpretation of crystal behaviours under variable electric fields.

\section{Computational models} \label{sec:Computational models}
\subsection{Guided Algorithm (GA)}\label{sec:model-a}
We used a guided error reduction (ER) and relaxed averaged alternating reflections (RAAR) algorithm \cite{Luke2005}, in conjunction with the shrinkwrap (SW) approach \cite{Marchesini2003}, to reconstruct BTO nanoparticles. The SW algorithm utilised a Gaussian fill function. To ensure uniqueness and reliability of the reconstruction, we incorporated multiple random starts in a guided algorithm (GA) framework that leveraged genetic breeding techniques \cite{Chen2007}. For the GA, we utilised a population of 100 with 5 generations of breeding. In each generation, the “\texttt{avg-half}" method was used, which computes the average across the better half of the population \cite{Ulvestad2017}. The breeding mode “\texttt{avg\_ab}" combines the best population iterate (“a"), determined by GA metrics, with the current population iterate (“b"), using a geometric mean \cite{Clark2015}.

To reduce noise, the diffraction pattern data were binned by a factor of two along each axis of the detector plane. The SW threshold was varied between 12\% and 15\%, consistently yielding reconstructions with Chi square ($\chi^2$) values (Eq. \ref{eqn:chisq}) between 1.6\% and 1.8\%. The final model was reconstructed using a 12\% SW threshold, achieving a $\chi^2$ of 1.62\%.

\subsection{\textbf{Phase Transfer Algorithm} (PTA)}\label{sec:model-b}
To visualise the structural changes in real space, we need to invert the diffraction patterns to images. Previously it has been effective to use a Difference Fourier Algorithm (DFA) \cite{Henderson1971, Watari2011}. The DFA is based on the idea that each missing phase of the diffraction pattern has contributions from every point in the real space object; hence if there is a small local change of structure, the diffraction phase is almost unchanged and can be transferred to the measured diffraction amplitude. When we calculated real-space images by DFA, we found there were no localised features in the images, so concluded it was insensitive to the kind of domain-wall motions expected. 

We then employed a Phase Transfer Algorithm (PTA) instead: a single reconstruction was carried out and the diffraction phases were transferred to the data  measured at the other voltages. Like for the DFA, this served to reduce inconsistencies which might have arisen from noise propagation effects resulting in spurious differences between separate reconstructions. We note that these are phase domain structures, strong real-space phase objects, which are hard to reconstruct reliably. The PTA avoids the small variations between the different reconstructions (from random starts) by using just one solution to phase the structural differences. 

The comparison of diffraction patterns, both between measurements and with reconstructed models employs metrics such as the Chi square and Pearson correlation coefficient (PCC) as defined below:
\begin{equation}
\chi^2 = \frac{\sum (A - B)^2}{\sqrt{\sum A^2 \cdot \sum B^2}}
\label{eqn:chisq}
\end{equation}
\begin{equation}
\text{r} = \frac{\sum (A - \overline{A})(B - \overline{B})}{\sqrt{\sum (A - \overline{A})^2 \cdot \sum (B - \overline{B})^2}}
\label{eqn:pcc}
\end{equation}
where A and B are data arrays of model diffraction pattern and experimental data.

The PTA approach we used involves employing a Genetic Algorithm (GA) in MATLAB to reconstruct the merged diffraction patterns of the crystal at different applied electric fields. Instead of starting from a random phase, we used the single merged 0V reconstruction, described above, as the initial guess for subsequent voltage scans. This ensured consistent results and avoided reproducibility issues. For each voltage scan, guided ER and RAAR algorithms introduced in Section \ref{sec:model-a} were performed until the algorithm converged, accurately capturing the structural evolution of the BTO sample under different voltage levels.  The resulting reconstructions yielded $\chi^2$ values of 7.54\% (10V), 4.65\% (5V), 1.42\% (0V), 3.16\% (-5V), and 3.66\% (-10V).

\subsection{c/a ratio from domain phases}\label{sec:c/a ratio}
Our side-by-side model from Figure \ref{BTO rec} (a) and (b) predicts a shift of domain walls under applied electric fields to enlarge the domains polarised parallel to the field. This process should not change the unit cell structure, meaning a constant ratio between the lattice parameter. The $c/a$ ratio of the lattice parameter follows the equation below \cite{Suzana2023}:
\begin{equation}
\frac{c}{a} = \sqrt{1 + \frac{a}{2\sqrt{2}\pi} \left( \frac{\Delta \phi_1}{\Delta x_1} - \frac{\Delta \phi_2}{\Delta x_2} \right)}
\label{eqn:c/a ratio}
\end{equation}
where $\Delta \phi_1$ and $\Delta \phi_2$ are the positive and negative slope of the phase; $\Delta x_1$ and $\Delta x_2$ are the thickness of the domain corresponding to the phase changes. The derivation of Eq. \ref{eqn:c/a ratio} is provided in the Appendix A. The lattice parameter of BTO 240 °C at room temperature are: a = 3.9942 \AA, c = 4.0305 \AA \cite{Suzana2023, Billinge2007}. The c/a ratio is therefore 1.0091.
\begin{figure}[ht] 
    \centering 
    \includegraphics[scale=0.128]{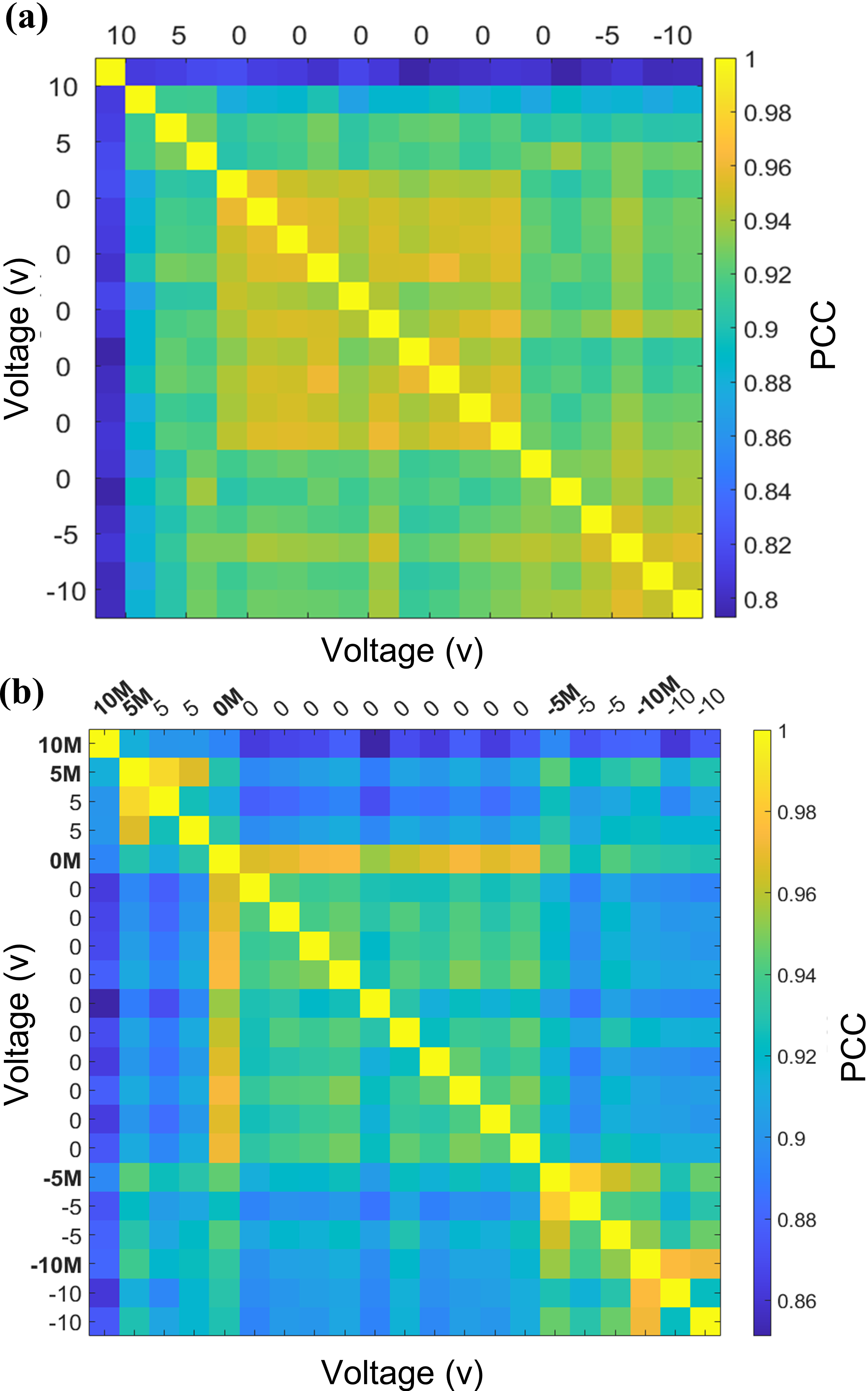}
    \caption[The caption that goes into the list of figures]{(a) Comparative analysis of voltage scans for a single BTO nanoparticle using the Pearson cross correlation (PCC) of the full 3D BCDI diffraction patterns to identify changes with applied voltage. (b) PCC analysis incorporating merged scans, denoted by "M". The corresponding voltage levels are labelled on the axes.
 }
    \label{BTO heatmap}
\end{figure}

\begin{figure*}[ht] 
    \centering 
    \includegraphics[scale=0.125]{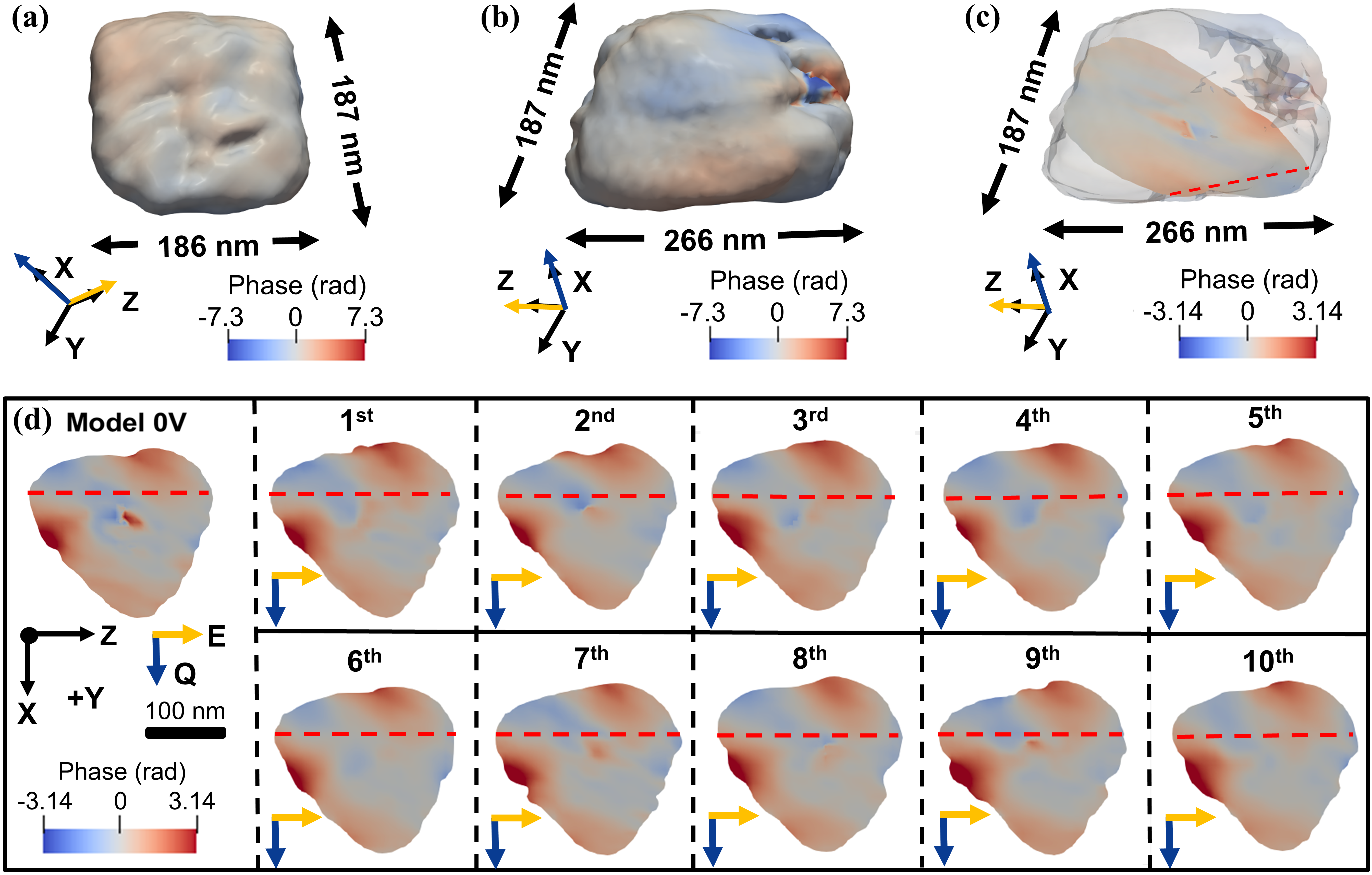}
    \caption[The caption that goes into the list of figures]{(a-b) 3D reconstruction of a BTO nanoparticle with no applied voltage, measured at 186 nm × 187 nm × 266 nm. The phase range is -7.3 to 7.3 rad. (c) Cross-section on the XZ (+Y) plane showing phase variations as colour changes indicative of the [110] lattice displacement. (d) Cross-sectional view of the BTO nanocrystal, reconstructed from ten diffraction patterns at 0 V (‘Model 0V’). Using PTA, ten slices were sequentially reconstructed based on ‘Model 0V’. A 100 nm scale bar is included with the blue $Q$ vector and the orange $E$-field aligned along the x and z-axis, respectively.}
 
    \label{BTO 12 0v}
\end{figure*}

\begin{figure}[b] 
    \centering 
    \includegraphics[scale=0.11]{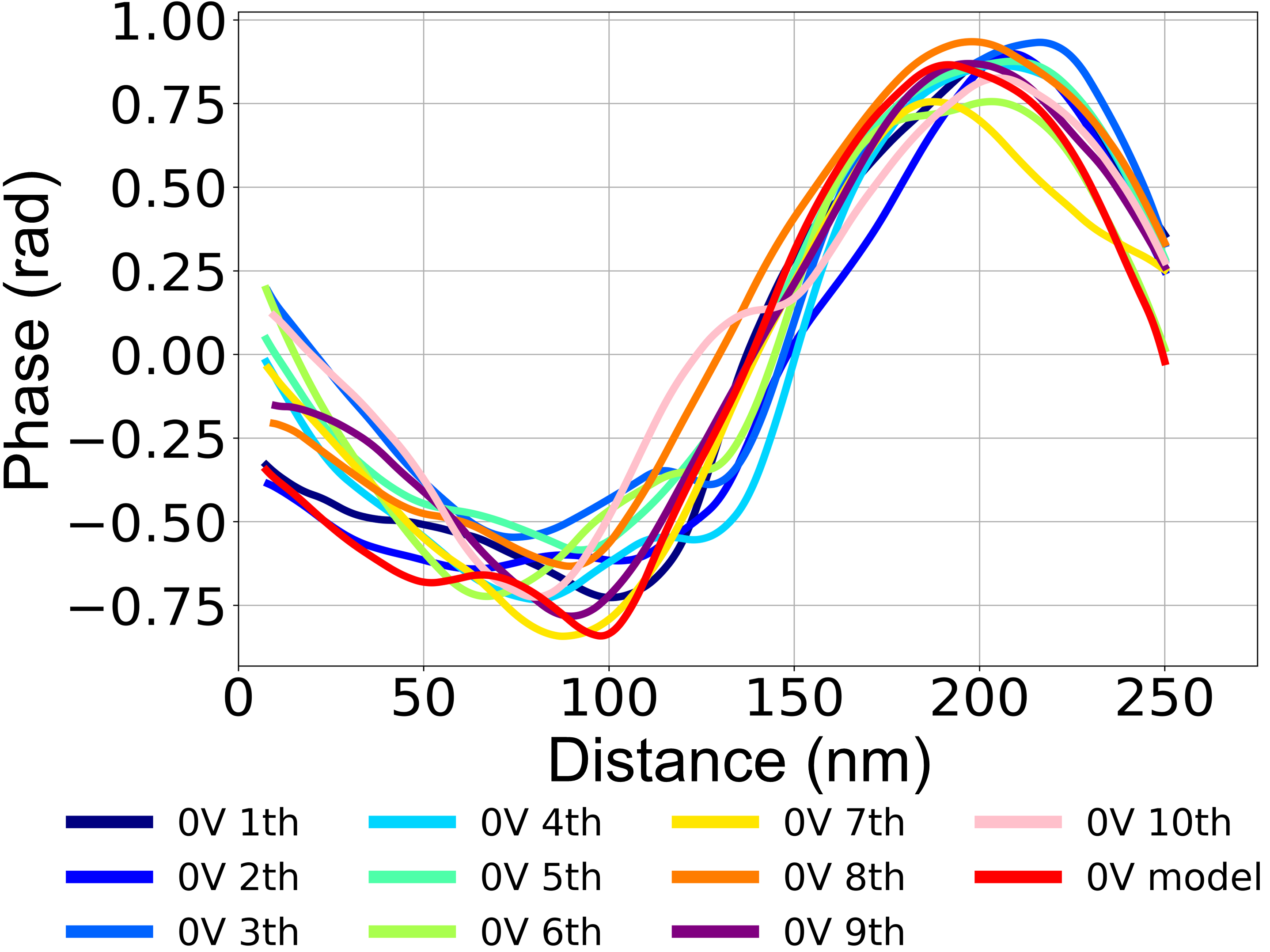}
    \caption[The caption that goes into the list of figures]{Phase line profiles along the red dashed line across the nanocrystal in Figure \ref{BTO 12 0v} (d), from slices 1-10 and 0V model.}
 
    \label{BTO 12 0v phase}
\end{figure}

\begin{figure*}[ht] 
    \centering 
    \includegraphics[scale=0.205]{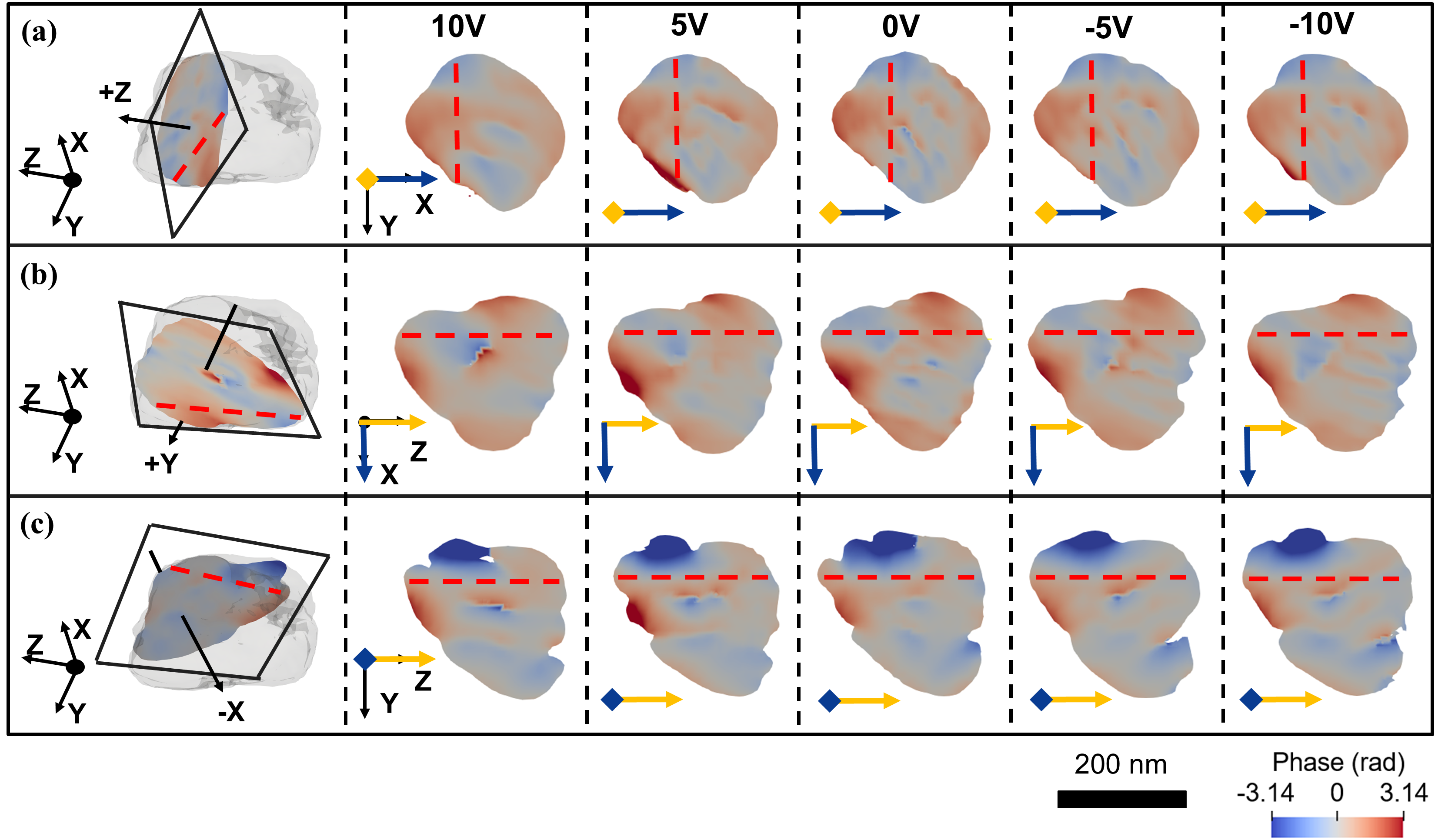}
    \caption{Panels (a-c) show isovolumes and 2D slices of a BTO nanoparticle at various voltage levels across the XY ($+$X), XZ ($+$Y), and YZ ($-$X) planes in laboratory cartesian coordinates. The E-field (orange arrow) aligns with the z-axis, and the blue $\boldsymbol{Q}$ vectors point roughly in the x-direction. Dashed red phase lines are perpendicular to $\boldsymbol{Q}$. }
    \label{GA rec ER 50 iterations}
\end{figure*}

\section{Results} \label{sec:Simulation}
The result is shown in Figure \ref{BTO heatmap} (a) which compares all the measured diffraction patterns under varying electric field strengths using PCC (Eq. \ref{eqn:pcc}) method. Each scan is compared with the rest of the voltage scans, presented as a 20 by 20 heat colour map. Results are set in an ordered sequence from 10 V to -10 V. 

At 0 volts, PCC values were high (above 0.94), indicating a good consistency of the measurement. These minor differences observed can be attributed to systematic errors inherent in BCDI measurements, such as beam stability or sample positioning variations. However, at $\pm 10$ and $\pm 5$ volts, there was a notable shift: PCC values decreased to as low as 0.79. This significant change in diffraction patterns suggests a voltage-induced alteration in the BTO crystal domain structure. Additionally, we removed three outliers, one in the initial 10 V scan and discrepancies in the last two 0 volt scans. These might arise due to sample damage or dielectric breakdown from the high voltages.

After the outliers were excluded to ensure consistency, the scans with identical voltage levels were merged together by cross correlation using sub-pixel shifts, enhanced the signal-to-noise ratio of the data. The PCC heat map in Figure \ref{BTO heatmap} (b) presents the merged scan data, showing clearer differentiation between voltage regions. This improved our ability to observe the subtle structural changes within the BTO nano-crystal under varied electric fields.

The merged scan labelled ‘0M’ is the sum of the ten 0 V diffraction pattern with consistent PCC values from 0.94 to 0.98. This merged scan was subjected to phase reconstruction as discussed in section III A. Its solution becomes the reference structure input for the PTA. Figure \ref{BTO 12 0v} (a) and (b) show the 3D reconstruction of the diffraction pattern of the merged 0 V scan. The sample has a rectangular shape with dimensions of $186  \pm 5 \, \text{nm}$ in length and width, and $266 \pm 5 \, \text{nm}$ in height. The nanocrystal size varies depending on the SW threshold, ranging from 12\% to 15\%.  A cross-section in the XZ plane is shown in panel (c), where the positive red phase indicates regions of the crystal displaced in the direction of the $\boldsymbol{Q}$ vector, while the negative blue represents displacement in the opposite direction. The same slice in panel (d), labelled “Model 0V", further highlights strain patterns at the central region, suggesting the presence of a dislocation loop. This dislocation serves as a fiducial marker when comparing reconstruction slices of the same crystal under different voltages. This slice was selected because it aligns with the applied voltage along the z-axis, where domain wall motion is expected. This allows evaluation of PTA consistency, as the phase along the red dashed line should remain uniform at 0V.

To assess the consistency of our PTA, we analysed the phase reconstruction patterns of 10 separate zero voltage scans. Figure \ref{BTO 12 0v} (d) displays the zero voltage phase maps obtained sequentially from the experiment, demonstrating similar patterns. Phase values extracted along a line from these maps, chosen to cross several domains, were used to evaluate domain size consistency across reconstructions. 

Figure \ref{BTO 12 0v phase} shows a close similarity with the triangular phase profiles seen by Suzana et al. \cite{Suzana2023}. Minor variations in slope are observed between 0 and 100 nm, likely due to noise. These deviations are more pronounced at the edges of the nanocrystal. The phase profile in Figure \ref{BTO 12 0v phase}, corresponding to the red dashed line in the schematic (Figure \ref{BTO rec} (a)), shows maxima and minima that indicate domain boundaries, aligning with the blue lines in Figure \ref{BTO rec} (a). The distance between these extrema represents domain thickness, measured at approximately 100 nm. The \( c/a \) ratio, derived from phase changes over these domain distances, yields a mean of 1.00052 with a standard deviation of 5.9 × 10\textsuperscript{-5}, closely matching the reference ‘Model 0V’ value of 1.00056. This strong agreement validates the reliability of our comparison algorithm for zero voltage conditions. When a voltage is applied, domain wall motion is expected to shift these extrema horizontally, while maintaining a consistent linear slope, as will be discussed in the following section.

\begin{figure}[ht] 
    \centering 
    \includegraphics[scale=0.17]{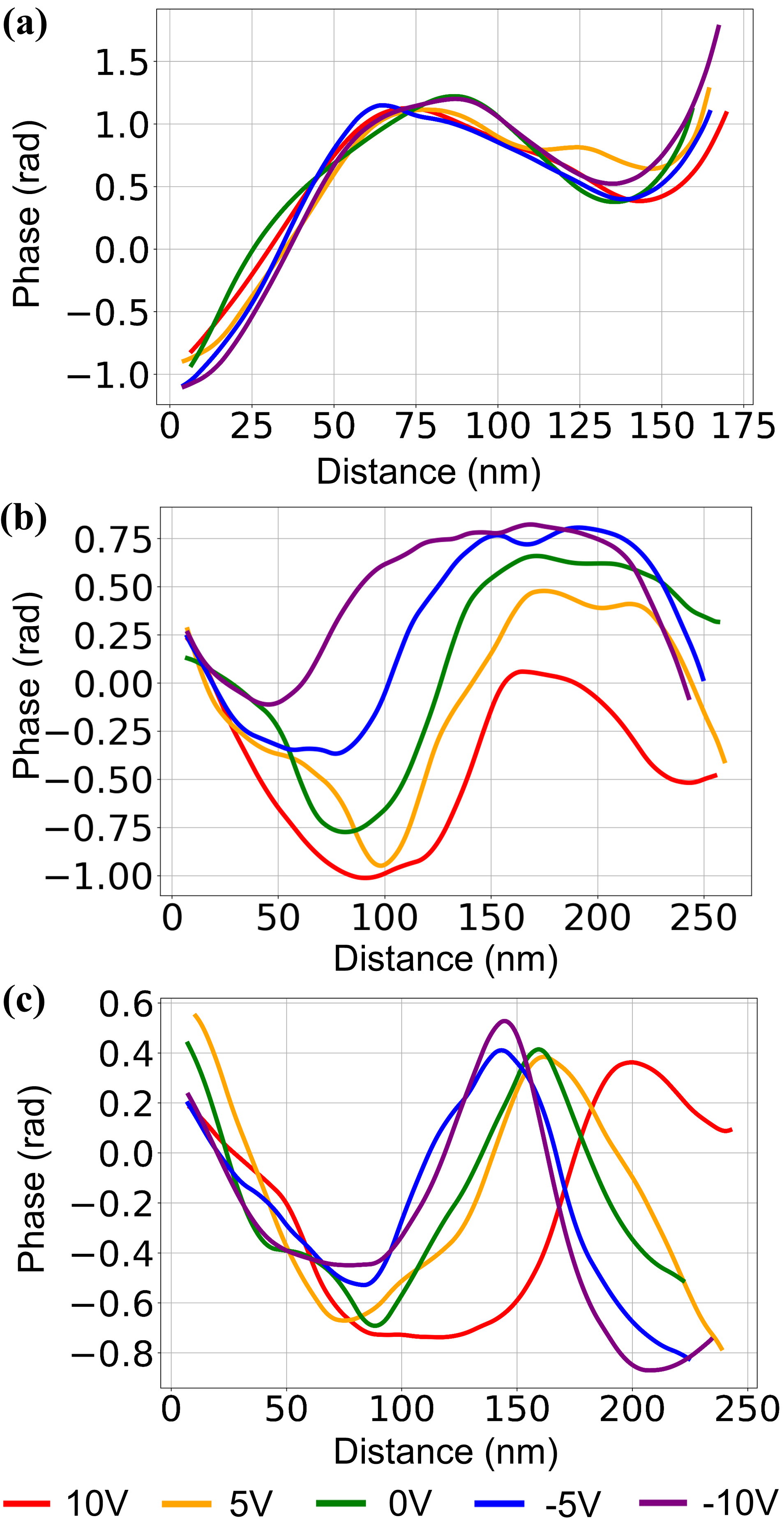}
    \caption{Panels (a–c) show phase line profiles along the dashed red line at different voltages from Figure \ref{GA rec ER 50 iterations}, with domain distances in nanometers. The line is perpendicular to the $\boldsymbol{Q}$ vector, highlighting regions of the largest phase shifts. Panel (a) represents the XY ($+$Z) plane, (b) the XZ ($+$Y) plane, and (c) the YZ ($-$X) plane.}
    \label{GA rec phase plot}
\end{figure}

\section{Phase Profile} \label{sec:Analysis}
Figure \ref{GA rec ER 50 iterations} (a-c) displays domain structures within the same BTO nanoparticle under different voltages. The $Q$ vector points along the [110] direction. The dashed red lines indicate the locations of the extracted phase plots, perpendicular to the $Q$ vector. These choices are aimed at determining whether the domain expansion is driven by the applied electric field. The reconstructed nanocrystals at each voltage are aligned using the center of mass (COM) method to ensure consistent comparisons across voltages. As mentioned previously, slices were aligned on a common central dislocation loop used as a fiducial marker.

In the $-$X view of the YZ plane, a triangle-shaped wave of in-plane shifts is seen to respond to the E-field. Similarly, the $+$Y view of the XZ plane shows migrated domain stripes due to out-of-plane shifts with increasing E-field. However, the $+$Z view of the XY plane shows no significant response. This suggestion of orientation-dependent domain behaviour is discussed below. The domain migration occurs predominantly along the Z-axis, which is further analysed in the phase plots below.

Figure \ref{GA rec phase plot} (a) illustrates the phase plot in the $+Z$ plane, where the domains remain stable, with a mean c/a ratio of 1.00081 (from slope, see Appendix A) and a standard deviation of 4.2 × 10\textsuperscript{-5}. 

In contrast, Figure \ref{GA rec phase plot} (b) highlights the most pronounced domain size changes and migration in a slice of the XZ plane. Under applied voltages, the domains extend up to 135 nm deep from the particle surface along the z-axis. At $+$5 V and $+$10 V, the phase-positive domain contracts by 25 nm and 35 nm, respectively, whereas it expands by 20 nm and 100 nm under $-$5 V and $-$10 V. The mean c/a ratio is calculated as 1.00078 with a standard deviation of 6.3 × 10\textsuperscript{-5}.

Figure \ref{GA rec phase plot} (c) also illustrates significant motions of domain walls in a YZ plane slice. At $+$10 V, the phase-positive domain migrates from 160 nm to 200 nm along the z-direction of the slice, while it shifts back to 145 nm at $-$5 V and $-$10 V. Likewise, the phase-negative domain migrates from 85 nm to 145 nm at $+$10 V and reverses by 10 nm at $-$5 V and $-$10 V. Migration effects are less pronounced at $+$5 V. The mean c/a ratio in this plane is 1.00066, with a standard deviation of 8.3 × 10\textsuperscript{-5}.

\section{Discussion} \label{sec:conclusions}
Our study offers new insights into the response of BTO nanocrystals under applied electric fields, employing BCDI to capture real-time domain behaviour at the nanoscale. With the electric field oriented approximately along the z-axis and the $Q$ vector positioned around 12 degrees off the x-axis, our results show a direct correlation between electric field strength and domain wall migration. The diffraction patterns in Figure \ref{BTO heatmap} (a) demonstrate a clear dielectric response to the external field, while the phase profiles in Figure \ref{GA rec phase plot} reveal substantial domain wall shifts and expansions. Notably, the consistent c/a ratios (ranging from 1.00066 to 1.00081) across all planes confirm the structural stability of the material, even under dynamic conditions.

BCDI real-space reconstructions in Figure \ref{GA rec phase plot} (b-c) clearly illustrate pronounced domain expansions and shifts along the z-axis, particularly in the XZ (+Y) and YZ (-X) plane slices. This behaviour can be attributed to the alignment of the $Q$ vector perpendicular to these planes, and phase lines parallel to the E-field, which maximises the observed effects. Conversely, no significant domain wall motion was detected when phase lines were perpendicular to the E-field, indicating an orientation-dependent response. The most significant domain wall motion occurred perpendicular to the $Q$ vector, driven by the E-field along the z-axis.

We selected the red dashed lines in the phase plots (Figure \ref{GA rec ER 50 iterations} b,c and Figure \ref{GA rec phase plot} b,c) to capture the regions of most significant phase shifts, which occurred predominantly at the surface of the BTO nanocrystals. This can be attributed to the nearly strain-free nature of the open surfaces, allowing domain walls to move more freely compared with those in the interior, where domains are constrained by strain and more tightly locked. This higher degree of freedom for domain motion at the surface may explain why nano-sized BTO exhibits a higher dielectric constant than bulk materials, a phenomenon first discovered by Wada et al. \cite{Wada2003}. In contrast, the strain in bulk materials restricts the motion of domain walls, limiting this mechanism of dielectric response.

While the domain shifts remained largely reversible across various electric field strengths, a notable loss of reversibility was observed at the highest voltage ($+$10 V). As depicted in Figure \ref{BTO heatmap} (a), the diffraction patterns at 0V exhibited permanent changes after applying $+$10 V. This suggests that $+$10 V represents a critical threshold, beyond which the influence of the electric field on the crystal structure becomes pronounced, showing plastic deformation. At such extreme voltages, the mechanisms driving domain migration may shift from purely reversible electric field effects to irreversible and might be a structural explanation of dielectric break down. Some other regions with steep slopes (high c/a ratio) in the phase plots further indicate that additional factors, such as intrinsic defects, may contribute to the limited reversibility of domain behaviour.

Our study confirms the side-by-side model of domain wall migration under external E-field and highlights critical thresholds for significant structural changes in BTO nanocrystals. Furthermore, this work provide insight into explain the enhanced dielectric response of nano-sized materials. These findings hold significant implications for future applications in nanoelectronic devices and materials engineering, where controlling domain behaviour and dielectric properties at the nanoscale is crucial.

\section*{Acknowledgements} \label{sec:acknowledgements}
We acknowledge the ESRF for the provision of synchrotron radiation facilities under proposal number HC-5370 utilising beamline ID01. The DOI for the data measured at ESRF is 10.15151/ESRF-ES-1199261421. Work at Brookhaven National Laboratory was supported by the U.S. Department of Energy, Office of Science, Office of Basic Energy Sciences, under Contract No. DESC0012704.

\begin{figure*}[ht] 
    \centering 
    \includegraphics[scale=0.5]{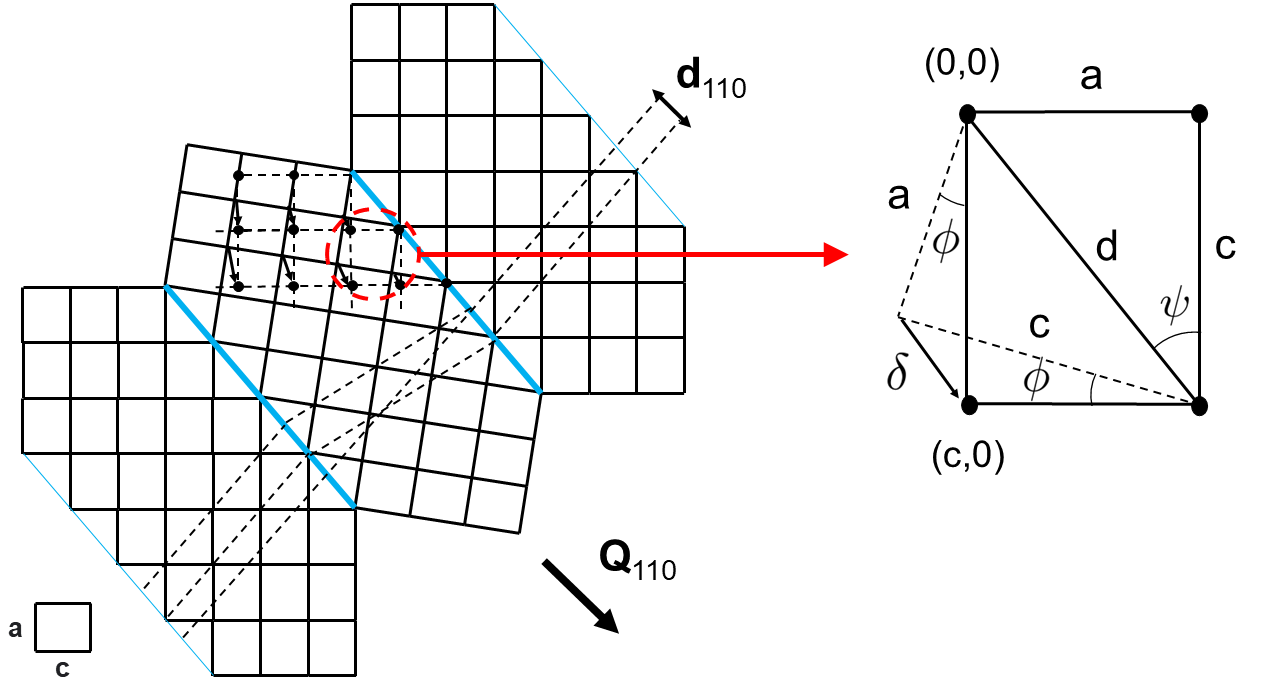}
    \caption{An exaggerated side-by-side model, provided for detailed calculation of the c/a ratio from phase domains.}
    \label{Appendix A}
\end{figure*}
\newpage

\section*{Appendix A}
\appendix

The trigonometric relations can be written as:
\begin{equation}
\begin{split}
\tan \psi &= \frac{a}{c} \\
\sin \psi &= \frac{a}{d} \\
\cos \psi &= \frac{c}{d}\\
\end{split}
\end{equation}

The shift $\delta$ is therefore given by:
\begin{equation}
\begin{split}
 \delta &= (c - a \cos \phi, a \sin \phi)\\
&= (c - a \sin 2\psi, a \cos 2\psi)
\end{split}
\end{equation}

Starting with the definition of \( |\delta|^2 \) and proceeding with the simplifications using Eq. \ref{eqn:eq7}:
\begin{equation}
\begin{split}
\sin(2\psi) &= 2\sin(\psi)\cos(\psi) = \frac{2ac}{d^2} \\
\cos(2\psi) &= \cos^2(\psi) - \sin^2(\psi) = \frac{(c^2-a^2)}{d^2} \\
\end{split}
\label{eqn:eq7}
\end{equation}

\begin{equation}
\begin{split}
|\delta|^2 &= (c - a\sin(2\psi))^2 + (a\cos(2\psi))^2 \\
       &= c^2 - 2ac\sin(2\psi) + a^2\sin^2(2\psi) + a^2\cos^2(2\psi) \\
       &= c^2 \left( 1 - \frac{2a^2}{d^2} \right)^2 + \frac{a^2}{d^4} (c^2-a^2)^2 \\
       &= \frac{1}{d^4} (a^6 + c^6 -a^4c^2 - a^2c^4) \\
       &= \frac{a^4(a^2-c^2)^2 + c^4(c^2-a^2)}{(a^2+c^2)^2} \\
       &= \frac{(a^2-c^2)^2}{(a^2+c^2)} 
\end{split}
\end{equation}
Taking the positive root,
\begin{equation}
\begin{split}
|\delta| &= \frac{(c/a)^2-1}{\sqrt{1+(c/a)^2}} \cdot a \\
\end{split}
\end{equation}
The resulting \( |\delta| \) can be used to calculate the displacement \( u \) over a lateral distance \( x \):
\begin{equation}
\begin{split}
u &= |\delta| \frac{x}{a/\sqrt{2}} \\
  &= \sqrt{2} x \left( \frac{(c/a)^2-1}{\sqrt{1+(c/a)^2}} \right) \\
  &\approx x \left( (c/a)^2-1 \right) \\
\end{split}
\end{equation}

The phase \( \Phi \) introduced by the displacement \( u \) is:
\begin{equation}
\begin{split}
\Phi &= Q \cdot u \\
     &= \frac{2\pi}{a/\sqrt{2}} \cdot u \\
     &= 2\pi \sqrt{2} \left( (c/a)^2-1 \right) \cdot \frac{x}{a} \\
\end{split}
\end{equation}
Rearrange to get c/a ratio:
\begin{equation}
\begin{split}
\frac{c}{a} &= \sqrt{1 + \frac{a}{2\sqrt{2}\pi} \left( \frac{\Phi}{x} \right)}\\
            &= \sqrt{1 + \frac{a}{2\sqrt{2}\pi} \left( \frac{\Delta \phi_1}{\Delta x_1} - \frac{\Delta \phi_2}{\Delta x_2} \right)}\\
\end{split}
\end{equation}
Where $\Delta \phi_1$ and $\Delta \phi_2$ are the positive and negative slope of the phase; $\Delta x_1$ and $\Delta x_2$ are the thickness of the domain corresponding to the phase changes.

\end{document}